\documentclass[preprint,5p,times]{elsarticle}

\usepackage{graphics}

\usepackage{amssymb}

 \usepackage{lineno}

\usepackage{multirow} 
\usepackage{hhline}

\usepackage[utf8]{inputenc}
\usepackage{fancyvrb}

\usepackage{hyperref}

\journal{Fusion Engineering and Design}

\begin{document}

\begin{frontmatter}



\title{Integrated Data Acquisition, Storage, Retrieval and Processing Using the COMPASS DataBase (CDB)}

 \author[label1]{J.~Urban}
 \author[label1]{J.~Pipek}
 \author[label1]{M.~Hron} \author[label1,label2]{F.~Janky} 
 \author[label1,label2]{R.~Pap\v{r}ok}
 \author[label1,label2]{M.~Peterka}
 \author[label3]{A.S.~Duarte} 
 \address[label1]{Institute of Plasma Physics AS CR, v.v.i., Za~Slovankou 3, 182 00 Praha 8, Czech Republic}
 \address[label2]{Department of Surface and Plasma Science, Faculty of Mathematics and Physics, Charles University in Prague, V Hole\v{s}ovi\v{c}k\'ach 2, 180~00 Praha 8, Czech Republic}
 \address[label3]{Instituto de Plasmas e Fusão Nuclear, Instituto Superior Técnico, Universidade Técnica de Lisboa, 1049-001 Lisboa, Portugal}

\begin{abstract} 
We present a complex data handling system
for the COMPASS tokamak, operated by IPP ASCR Prague, Czech Republic \cite{compass2006}.
The system, called CDB (Compass DataBase),
integrates different data sources as an assortment of data acquisition hardware
and software from different vendors is used. Based on widely available open
source technologies wherever possible, CDB is vendor and platform independent
and it can be easily scaled and distributed.
The data is directly stored and retrieved using a standard NAS (Network Attached
Storage), hence independent of the particular technology; the description of the
data (the metadata) is recorded in a relational database. Database structure is
general and enables the inclusion of multi-dimensional data signals in multiple
revisions (no data is overwritten).
This design is inherently distributed as the work is off-loaded to the clients.
Both NAS and database can be implemented and
optimized for fast local access as well as secure remote access.
CDB is implemented in Python language; bindings for
Java, C/C++, IDL and Matlab are provided.
Independent data acquisitions systems as well as nodes managed by FireSignal \cite{FireSignalRef}
are all integrated using CDB.
An automated data post-processing server is a part of CDB. Based on dependency rules,
the server executes, in parallel if possible, prescribed post-processing tasks.

\end{abstract}


\end{frontmatter}


\section{Introduction}
\label{sec:intro}

With the increasing volume and complexity of diagnostics and synthetic data needed for tokamaks or
other pulsed experimental devices, the demands on the data storage system are becoming very
challenging. Present technologies are often difficult to scale, either due to the used technologies
or due to the internal architecture. Since the COMPASS tokamak (re)started its operation at IPP
Prague \cite{compass2006}, more and more issues related to data storage emerged. This fact,
together with the experience from other tokamaks, motivated the development of the \emph{COMPASS DataBase}
(CDB)---a system to be used to store and retrieve any (COMPASS) tokamak related numerical data.

This paper first discusses the fundamental goals and motives of CDB in Section \ref{sec:goals}. 
The CDB architecture is
described in Section \ref{sec:arch}. The core functionality is implemented in
Python with bindings for many different languages and interfaces, as described in Section 
\ref{sec:implement}. The integration of the various data acquisition systems
is addressed in Section \ref{sec:daq}. A recently implemented automatic post-processing
capability is described in Section \ref{sec:postproc}. 
Finally, Section \ref{sec:concl} gives our conclusions.


\section{Goals and motivation}
\label{sec:goals}

As already noted above, it was the re-installation of the COMPASS tokamak at IPP Prague that
provided the fundamental motivation for the development of a new data storage system. Within a close
collaboration with IPFN Lisbon, a new CODAC (Control, Data Access and Communication)
system \cite{atcacompass, COMPASSCODAC2013} has been built. In the beginning, the system consisted of ATCA data
acquisition hardware and FireSignal \cite{FireSignalRef}, which controlled the data acquisition
nodes. SDAS \cite{SDASRef} was used to access the data provided by FireSignal. This original design
was strongly focused on data acquisition as the data signals were identified by hardware identifiers
while the natural identification is the (measured) physical
quantity. Moreover, data are written to and read from a central server, which can thus become overloaded. 
This is the case of COMPASS as around 2~GB of diagnostic data are presently produced for a single
discharge; this number can become several times larger in the near future.
Moreover, the API was not very flexible and numerical data were stored in custom binary files.

\begin{figure*}
\centering   
\hfill{}
\includegraphics[width=13cm]{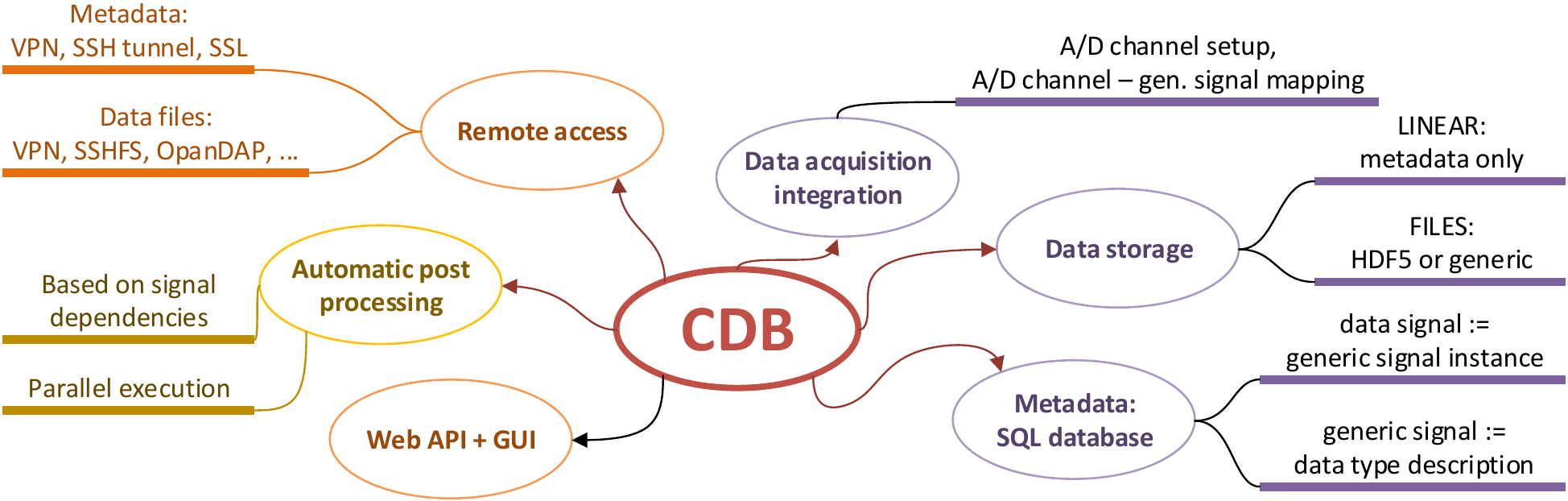}
\hfill{}
\caption{An overview diagram of CDB components.}
\label{fig:CDBmm}
\end{figure*}

\emph{COMPASS DataBase} (CDB) has been designed to enhance the capabilities and performance of COMPASS
CODAC, focusing mainly on the final storage system. CDB
provides a way to write data from data acquisition nodes in parallel, without a central
collection point. As such, the FireSignal central server can be used for nodes control only, while the 
data are written directly to CDB.
CDB is
easily scalable, portable and extensible. Although targeted to be used primarily on COMPASS, the aim
is to create a universal tool. The CDB architecture is depicted in Fig. \ref{fig:CDBmm} and
described in detail in the following.

\section{The architecture of CDB}
\label{sec:arch}

\subsection{Data and metadata}

Scientific data that we need to store are well structured (we know in advance the type, the
dimensionality, etc.) and can be decomposed into actual data (numbers) and metadata, which contain
the information about the data content (e.g., the physical quantity, units, etc.).

\begin{figure}
\centering   
\hfill{}
\includegraphics[width=8cm]{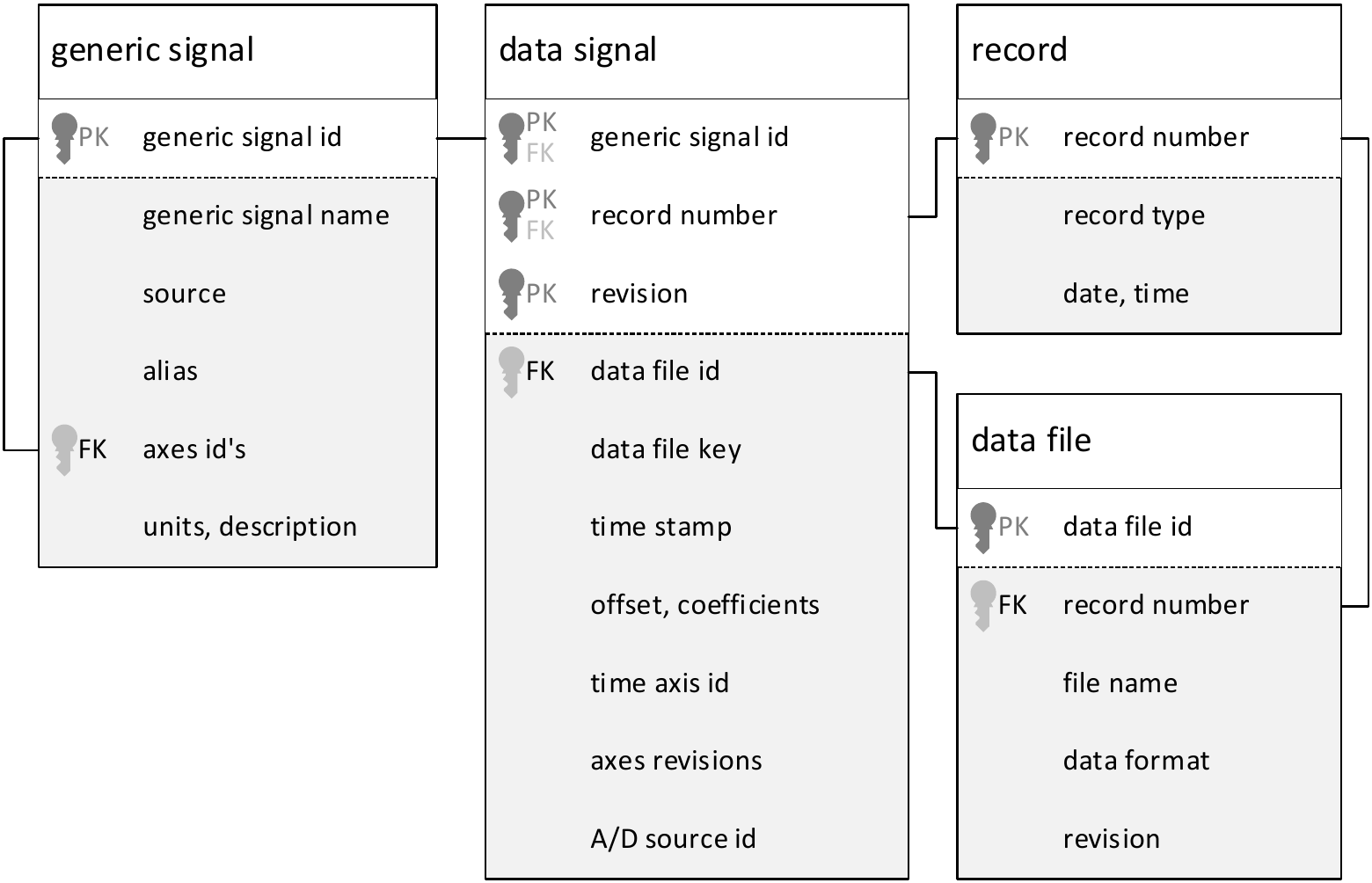}
\hfill{}
\caption{A simplified diagram of the CDB data model. We use an SQL-like graphics since the data 
structure is directly reflected in the SQL schema. Keys denote primary identifiers (primary keys 
in SQL), lines denote references (SQL foreign keys). Certain fields, e.g. axes id's, represent 
multiple fields in reality.}
\label{fig:datamodel}
\end{figure}

The data model of CDB, depicted in Fig. \ref{fig:datamodel}, is based on \emph{generic signals},
which contain the descriptions of possible data types that can be stored, and \emph{data signals},
which are instances (realizations) of generic signals and contain the \emph{metadata} of a particular 
dataset, including the \emph{data file}, in which the numerical data are stored.
Generic signal description also contains the axes, which are
generic signals as well. 

A single data signal as well as single data file always belong to a particular \emph{record number}
and can exist in one or more \emph{revisions}. The record number denotes either a particular
experimental discharge or a model (simulation) or a void (e.g. a trigger test) record.

Two types of data signals can be stored in CDB: FILE and LINEAR. FILE signals have the data in
data files while LINEAR signals are described by a linear function.
Linear transforms of FILE signals can also be stored, particularly for converting from data acquisition 
levels to physical units or for correcting the data, so that no new data file has to be created.

All metadata, i.e., generic and data signals, records, data file descriptions and more,
are stored in a relational database, particularly MySQL, although
a different database engine can be used. 
The numerical data are stored in files on a network attached storage (NAS). The primary file format is
HDF5 \cite{HDF5Ref}. 
There are many good reasons to employ HDF5 files on a NAS as the numerical data
storage back end. HDF5 is one of the most enhanced and wide spread file formats, which allows to
conveniently read and write almost any kind of numerical datasets, organized hierarchically in a
file. Very importantly, HDF5 API is extremely rich in its functionality and is available for all
relevant programming languages. 
Each HDF5 file can contain one or more CDB data signals. CDB also allows to store any other data type, 
however, without providing any compatibility to read the data. This feature can be convenient, for example, 
for storing whole custom data files of a simulation code (instead of storing it ``somewhere'', CDB stores 
the file and its metadata in a well defined place) or for storing encoded videos or images.

Since CDB uses conventional file access protocol, any NAS can be used to read and write the data
files. This is, in most cases, a big advantage as the storage can be tailored and optimized
independently of CDB itself. A cluster storage system is actually used for COMPASS
\cite{GlusterCOMPASS}. 
To increase the performance of writing data from FireSignal, we have an additional local cache on the
central server.

\subsection{Signal identifiers}

CDB \emph{generic signals} are uniquely identified either by their numeric id, by a combination of name and
data source or by an alias. In order to have a unified API for different
languages and for different (and extensible) identification schemas, CDB uses string identifiers.
We define \verb!gs_str_id! (generic signal string id) as a string that contains one of the unique
generic signal identifiers.

\emph{Data signals} are uniquely identified by a unique generic signal identifier (see above) in
combination with a record number and a revision. For this reason, we define a string identifier
\verb!str_id!:

\begin{Verbatim}[fontsize=\footnotesize,commandchars=\\\{\}]
str_id := 
<CDB:> gs_str_id <:record_number <:revision>> <[units]> |
FS | DAQ: channel_id <:record_number <:revision>> <[units]>
\textrm{where}
channel_id := computer_id/board_id/channel_id
\end{Verbatim}

Here CDB, FS and DAQ are schemas for signal identification. The default is the native CDB
schema while FS and DAQ are used for identification by data acquisition channels using FireSignal or
CDB native id's, respectively. The following \verb!str_id! examples refer to the same signal, 
the first one by its alias while the others by its DAQ and FireSignal id's:
\begin{enumerate}[1.]
\item \verb!I_plasma:4073:-1[default]!
\item \verb!DAQ:ATCA_1/9/13:-1!
\item \verb!FS:PCIE_ATCA_ADC_01/BOARD_9/CHANNEL_013:4073!
\end{enumerate}
String identifiers can be used in any programming language, hence simplifying and
unifying the bindings, and are easy to compose and parse. They can also be straightforwardly
extended for, e.g., multiple tokamaks (by adding the tokamak name as a prefix), different databases
or even indexing and slicing. 
An additional schema can be implemented if needed. In particular, an ITM CPO \cite{ImbeauxCPO} schema is planned to be 
implemented, which would allow to search signals by CPO field names.

\begin{figure}
\centering   
\hfill{}
\includegraphics[width=6cm]{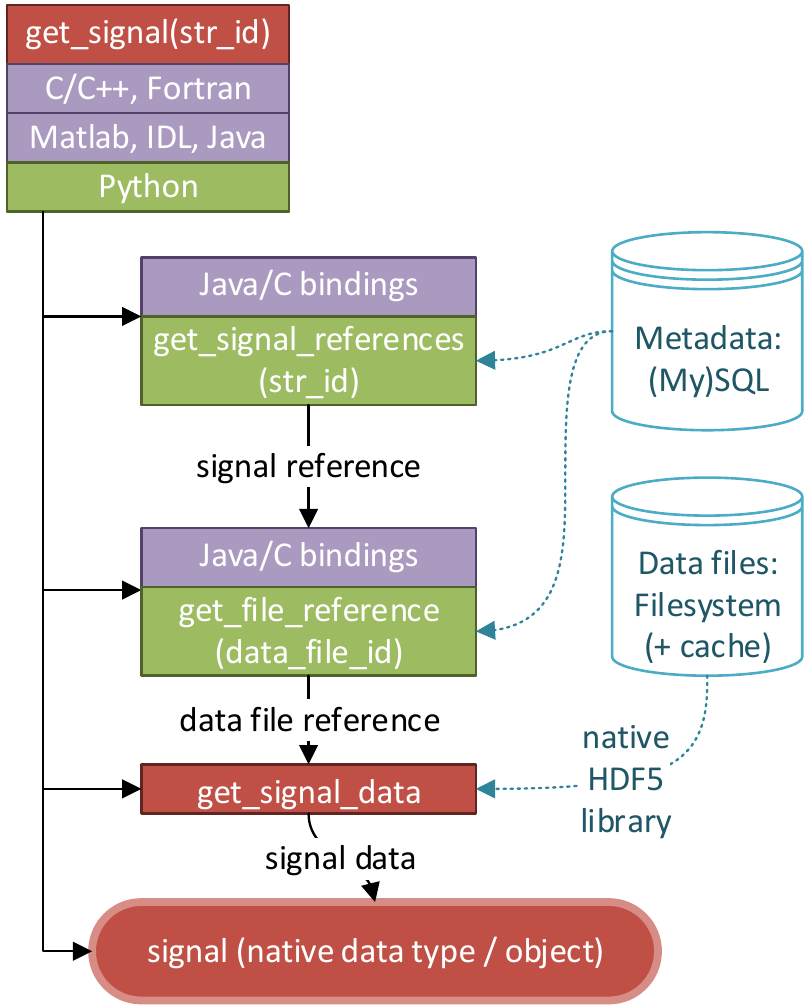}
\hfill{}
\caption{A schema of getting a signal from CDB. The starting point is the {\tt get\_signal} function in various languages. Core Python functions (green) are called via different APIs. Dotted lines show the use of back-end technologies. Finally, the data are assembled into a signal data type, which is native for the calling language.}
\label{fig:get_signal_schema}
\end{figure}

\subsection{Data safety and consistency, revisions}

When a new version of a signal is stored, e.g., in case of an error, a new revision is stored
instead of overwriting the existing data. As such, full history is saved and data consistency and
persistence is assured. When a data file is closed, CDB client changes its permissions to read only.
Moreover, the cache mechanism changes the owner to root and moves the files to the permanent storage, 
which can be read-only for clients.

\subsection{Data reading and storing}

CDB provides a simple data access along with a comprehensive set of functions for specific use
cases. This is particularly possible without too much effort due to the use of HDF5 files.
For simple data reading, function \verb!get_signal! \verb!(str_id)! reads a data signal including
all axes and numerical data, 
as shown in Fig. \ref{fig:get_signal_schema}.
This function returns either an object or a structure depending on the
programming language. Alternatively, we can request only signal reference(s) (i.e. the metadata) and use
the versatile HDF5 API to access the data. In our opinion, using HDF5 is extremely beneficial as we
save the tremendous effort that would be needed to re-implement similar functionality. Besides
simple dataset reading and writing, HDF5 natively supports reading portions of datasets
(hyperslabs), compound data types, point selection, extensible datasets, parallel access and even
more.

Storing data is naturally more complex. In general, the following steps are required to store data
into CDB. Function \verb!store_signal! is used in steps \ref{step:store1} -- \ref{step:store2}. Its
parameters are basically the fields from the data signal description (see Fig. \ref{fig:datamodel}).

\begin{enumerate}[1.]
\item Get the \emph{record number} or create a new record.
\item Get the \emph{generic signal reference} for the data to be stored.
\item\label{step:file1} Create a new data file---file path and name are obtained.
\item\label{step:file2} Write the numerical data into this file. Axes data can be written into the same file if convenient.
\item\label{step:file3} Close the file because CDB needs to know when writing to the file has finished.
\item\label{step:store1} Either store the \emph{axes data signals} or get their references if axes signals are already available. This must precede the next step because we must know the axes data signal revisions.
\item\label{step:store2} Store the main data signal.
\end{enumerate}
Steps \ref{step:file1} -- \ref{step:file3} are needed for FILE signals only.

Frequently, a simplified \texttt{put\_{}signal} function can be used, particularly when storing new
numerical data together with the axes. Another convenience function is \texttt{update\_{}signal},
which creates a new revision of a signal by changing only the metadata (offset, coefficients, time
axis, axes id's). Technically, this function replaces step \ref{step:file1} by getting the reference
of the signal to be updated, skips steps \ref{step:file2} -- \ref{step:store1} and finally stores
the signal using the new metadata.

\subsection{Remote access}

Remote access to CDB is provided by making accessible the data and metadata storage and configuring 
the CDB client accordingly by specifying the SQL server address and the data files root directory. 
There are no specific parts of the CDB code base dedicated to remote access. Available solutions, such 
as SSH, VPN or global file systems, can readily be used.

\subsection{Data acquisition set up}

CDB is capable to keep track of the relations between data acquisition channels and 
generic signals along with a text description of the configuration (e.g. the configuration file).
In other words, CDB knows what quantity is acquired by a particular A/D channel. 
This information can be very important since almost thousand channels are installed. 
Consequently, stored 
signals can be referenced either by generic signals or by A/D channel id's.

\section{Implementation details}
\label{sec:implement}

\subsection{Core functionality}

The core functionality is implemented in Python. Main reasons why Python is used are:
\begin{enumerate}[(a)]
\item Python is an open source, universal, widely used programming language.
\item Python is well equipped for interfacing to other languages.
\item Development in Python is rapid, which is supported by a huge set of available libraries. 
\item The native API can be readily used for data analysis and graphics, for which Python is well equipped.
\end{enumerate}
In fact, from the experience with CDB and other projects, we can state that Python is presently 
the most capable and versatile open source, high-level programming for scientific use.

\subsection{Bindings to other languages}

CDB bindings to other languages are conveniently provided using Jython and Cython. 
Jython is a Python implementation in Java; however, it can be equally well used to
execute Python from Java. Similarly, the primary purpose of Cython is writing C extensions in a Python-like 
language but it is also possible to create C API to Python functions.

Matlab and IDL are frequently used (commercial) products for scientific programming and data analysis. 
They can both call C functions by creating special wrappers; however, this capability is not very convenient, 
particularly in case of IDL, in which it is extremely user unfriendly. There are also problems with dynamically 
linked libraries. Finally, after implementing basic CDB bindings for Matlab and IDL using C and Cython, 
we rather decided to develop a Java-based interface. While Matlab runs Java natively, IDL is sometimes more 
peculiar. Nevertheless, the Python-Jython-Java solution proved to be most convenient and well maintainable 
for both Matlab and IDL, as well as for any other Java-based product.

An important feature of the CDB 
design is that the terminal languages use the bindings for metadata reading and writing only. Numerical 
data, i.e., HDF5 files in most cases, are read and written with native functions directly from/to the final 
file storage (NAS). This enhances the performance and relaxes the demands on servers. Although this solution 
implies that some parts of the code must be duplicated, the performance gain is more important.

If a programming language is required that cannot incorporate Python, Java or C/C++, CDB provides 
web services that map to CDB API and use JSON as the data format. Although the full 
potential of this interface has not been fully exploited yet, it clearly enables,
e.g., the creation of various web tools.

\section{Data acquisition integration with CDB}
\label{sec:daq}

On COMPASS, around 1000 data acquisition channels of different hardware characteristics as well as various modes 
of operation are installed. Among them are general-purpose multi-channel systems
with public hardware drivers and C-API (ATCA boards),
machines supporting remote network protocols (d-tAcq 216 or 196), hardware
requiring specific platform-dependent software 
controllable through a custom IPC protocol (infrared camera) or hardware controlled using applications written
in very specific programming languages. Some of them produce 1-dimensional data with linear time axes, 
others
produce multi-dimensional data with irregular samplings in time and space.
Although there is no single way to manage all these data-producing agents, most of them have been
easily integrated into CDB because of its flexibility and language independence. 

\subsection{FireSignal}
Most data acquisition systems 
are controlled
through FireSignal \cite{FireSignalRef}, a client-server network system. In its original form,
this system provides a \emph{central server} which is the hub of all communication---all other components
connect to it through CORBA and do not communicate with each other. A \emph{database controller} 
stores and retrieves data, which are provided by \emph{nodes}.
Users can connect via a graphical interface.

This architecture follows best design principles in that it conceptually divides the responsibility 
to subsystems and clearly separates the problem of data storage from the problem of control and
the data acquisition. However, in the case of COMPASS, this elegance of design brings about a few disadvantages,
most severe being the heavy load of the central server (most of which is just passing data around).
Therefore (and because of the change of the underlying database solution), we adopted a few changes
that violate the fundamental FireSignal architecture but which lead to higher efficiency
and employ CDB's potential:

\begin{enumerate}[(a)]
\item The database controller stores data into CDB (using Jython). Time axes, scaling coefficients
  and generic signals are automatically assigned in this step.
\item Data in the client GUI are read directly from CDB
  instead of being requested and received through central server.
\item There is a mechanism for the nodes to write data directly to CDB
  (using either Python, Cython or Jython, depending on the node implementation language)
  instead of sending them to the central server (and database controller). In such case, the nodes just
  inform the central server that they finished writing.
\item Multidimensional data and data without any time axis can be written from
  the nodes.
\end{enumerate}

The final goal is to write as much data as possible directly
from the nodes to CDB. For this purpose, we have developed C++ and Java libraries. 
Nevertheless, the implementation
of each node brings about specific complications.
In any case, the fall-back indirect mechanism of data storing through central server
will remain possible.

\subsection{Independent data acquisition sources}

Certain data acquisition systems do not easily fit in the FireSignal work flow, for which a special
node is either
very difficult (because of the API language or the operating system) or impractical (because of
the way they are configured or used between discharges). However,
if these systems know when to collect data, how
to collect data and where to write data, the data acquisition can be handled easily using triggers 
and the CDB interface.

\section{Automatic data post-processing}
\label{sec:postproc}

Numerous data post-processing tasks need to be executed after each tokamak discharge, ranging from a simple 
noise or drift removal to a magnetic equilibrium reconstruction. Moreover, it is extremely important 
to have the post-processed data consistent with the source data. If, for example, a source signal is 
updated (a new revision is created), the processing must be re-executed. As there seems to be no 
well suited solution available, we have decided to incorporate a post-processing capability into CDB.

Each CDB post-processing task has specified input and output signals. Tasks and their input 
and output signals form a directional graph, in which each task has edges oriented from the input signal nodes
and to the output signal nodes. The graph must be acyclic, i.e., iteration over 
multiple tasks is not allowed. In addition, each output signal must be created by exactly one task, otherwise 
dependencies are ambiguous. Post-processing tasks are stored in an SQL table and the prerequisites are 
checked whenever a new task is added.

Tasks are implemented as Python scripts, which either contain all the functionality or 
call external programs. The tasks are executed by traversing the tasks graph. \emph{IPython.parallel} 
is employed to execute the tasks in parallel. Running 
processes are periodically checked and new tasks are executed whenever all necessary inputs become available.
In addition, log entries are created, which allows to analyse the execution and also to execute the tasks 
only if newer inputs are available.

The post-processing is currently in a testing phase and launched manually. In the future, it 
will be triggered by storing a signal so that the post-processed signals will always be up to date. 

\section{A brief comparison to existing systems}

Two most similar products used in the tokamak community are IDAM \cite{IDAMRef} and MDSPlus \cite{MDSPlusHDF5}. 
A notable difference is that CDB does not enforce a specific data access protocol (although HDF5 is de facto native) while IDAM and MDSPlus feature a unified local and remote access.
CDB also directly exposes the meta data SQL server to clients, hence enabling custom queries. 
Composite signals (expressions) in IDAM (MDSPlus) can be replaced by the post-processing in CDB.

\section{Summary and conclusions}
\label{sec:concl}

A new data storage solution, the COMPASS DataBase (CDB), has been implemented for  
the COMPASS tokamak data. CDB integrates available and open technologies, such 
as HDF5 and MySQL, and provides a low-overhead, easily scalable and extensible solution for storing 
tokamak (or, generally, any pulsed device) experimental and simulation data. It has met the 
challenging needs of the COMPASS tokamak, particularly storing many data signals, which can originate
from many different data acquisition systems, in a convenient, transparent, consistent
and persistent form. 
CDB integrates easily with FireSignal as well as other systems, independent of their operating system 
or the API language.
Data post-processing, described by signal dependencies, is conveniently automated. 
CDB is available under MIT license at \url{http://bitbucket.org/compass-tokamak/cdb}.

\section*{Acknowledgement}
This work has been carried out within the framework of the Contract of
Association between EURATOM and the IPP.CR.  The views and opinions do not necessarily reflect those
of the European Commission.
The work of J. Urban was partially supported by Czech Science Foundation grant 13-38121P,
the work of M. Hron and F. Janky by Czech Science Foundation grant P205/11/2470,
the work at IPP AS CR by MSMT LM2011021.



\bibliographystyle{model1-num-names}
\bibliography{cdbiaea}






 
\end{document}